\documentclass[11pt,twoside]{article}

\usepackage{asp2006}
\usepackage{lscape}
\usepackage{amssymb}
\usepackage{graphicx}
\usepackage{aas_macros}

\begin{document}

\title{The angular power spectrum of the diffuse gamma-ray background as a probe of Galactic dark matter substructure}   
\author{Jennifer M. Siegal-Gaskins}   
\affil{Center for Cosmology and Astro-Particle Physics, The Ohio State University, Columbus, OH 43210}   

\begin{abstract} 
Dark matter annihilation in Galactic substructure produces diffuse gamma-ray emission of remarkably constant intensity across the sky, and in general this signal dominates over the smooth halo signal at angles greater than a few tens of degrees from the Galactic Center.  The large-scale isotropy of the emission from substructure suggests that it may be difficult to extract this Galactic dark matter signal from the extragalactic gamma-ray background.  I show that dark matter substructure induces characteristic small-scale anisotropies in the diffuse emission which may provide a robust means of distinguishing this component.  I discuss predictions for the angular power spectrum of the diffuse emission from various extragalactic source classes as well as from Galactic dark matter, and show that the energy dependence of the angular power spectrum of the total measured emission could be used to confidently identify a gamma-ray signal from Galactic dark matter.
\end{abstract}

\section{Introduction}
\label{sec:intro}
The observation of gamma-ray emission from the annihilation of dark matter particles could provide a means of indirectly detecting dark matter, constraining its properties, and mapping its spatial distribution.  The Large Area Telescope aboard the Fermi Gamma-ray Space Telescope \citep[\emph{Fermi},][]{atwood_abdo_ackermann_etal_09} will measure for the first time the all-sky gamma-ray emission at energies from a few tens of GeV to a few hundred GeV, an energy band which is especially promising for indirect dark matter searches.

\begin{figure}
\includegraphics[width=0.45\textwidth]{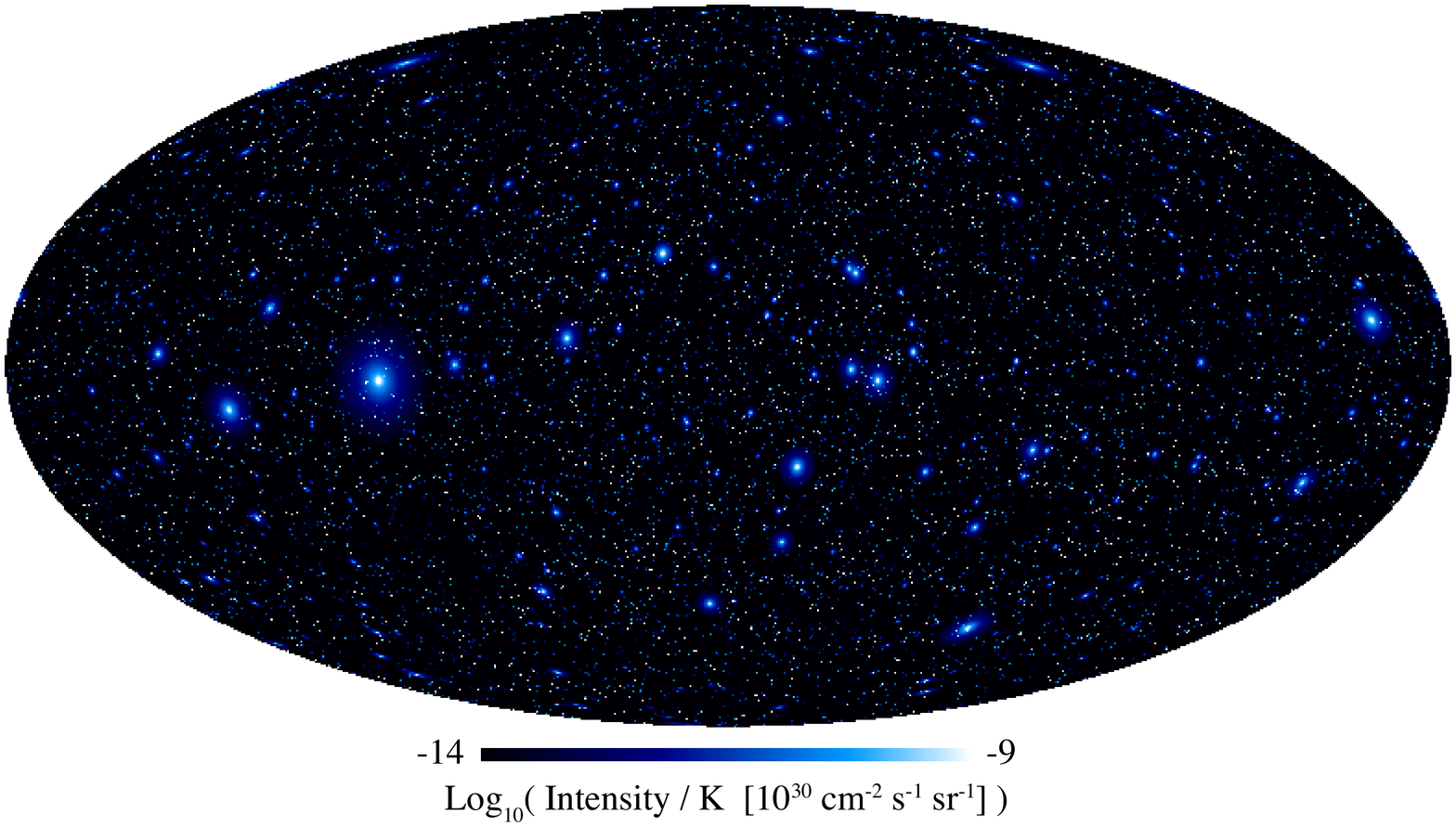}
\includegraphics[width=0.45\textwidth]{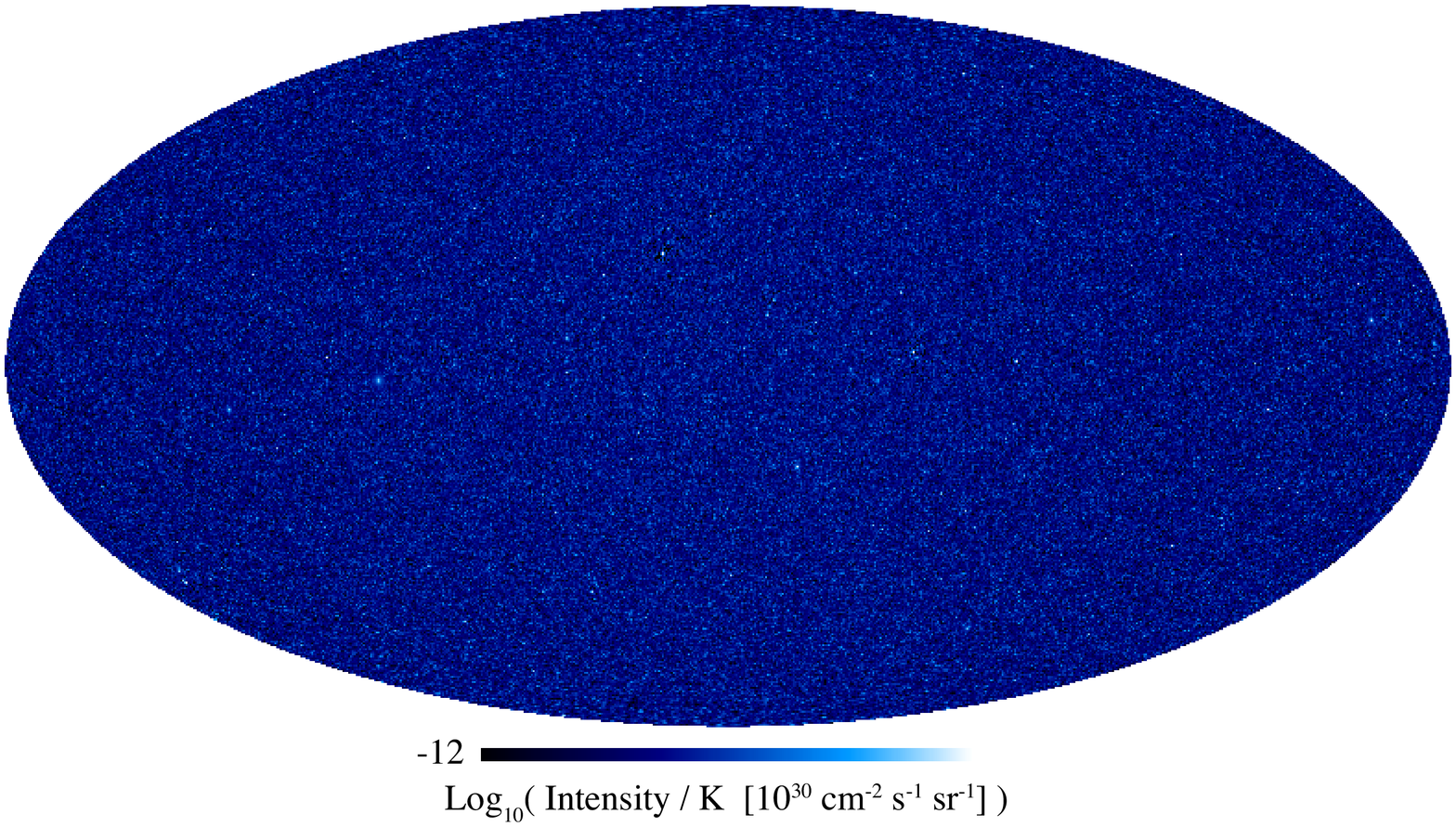}
\caption{Gamma-ray intensity from dark matter annihilation in Galactic substructure.  The right panel shows the same map as the left panel, but smoothed with a Gaussian beam of width 0.1$^{\circ}$, corresponding to {\it Fermi's} angular resolution above $\sim$ 10 GeV\@.  Figures are from \citet{siegal-gaskins_08}.
\label{fig:maps}}
\end{figure}

A generic prediction of cold dark matter models is the persistence of substructure within dark matter halos.  Depending on the details of the particle physics model, the smallest dark matter halos form in the early universe with masses of $10^{-12}$ -- $10^{-6}$ M$_{\odot}$ \citep[e.g.,][]{green_hofmann_schwarz_05,profumo_sigurdson_kamionkowski_06}, and many of these dense halos are expected to survive to the present day.  This prediction is consistent with the results of numerical simulations, which find an abundance of subhalos within a Galaxy-sized dark matter halo.  Simulations also find that the radial distribution of subhalos is less centrally concentrated than the smooth dark matter of the host halo \citep[e.g.,][]{gao_white_jenkins_etal_04,diemand_kuhlen_madau_07,kuhlen_diemand_madau_07,madau_diemand_kuhlen_08,springel_wang_vogelsberger_etal_08}, and consequently the angular distribution of Galactic subhalos on the sky as observed from our position appears fairly isotropic \citep{kuhlen_diemand_madau_08,siegal-gaskins_08,springel_white_frenk_etal_08}, as shown in the left panel of Figure~\ref{fig:maps}.

The rate of dark matter annihilation scales with the square of the density, so nearby regions of high density, such as the Galactic Center and subhalos, are natural targets for indirect searches.  While the Galactic Center is almost certainly the brightest point source of gamma-rays from dark matter, substantial astrophysical foregrounds make extracting a dark matter signal from the Galactic Center a formidable challenge.  Subhalos, on the other hand, are largely free of such foregrounds, and although most subhalos in the halo of the Milky Way are not likely to be resolved individually as point sources by \emph{Fermi} \citep{pieri_bertone_branchini_08,kuhlen_diemand_madau_08}, collectively substructure is expected to generate a large flux of diffuse gamma-rays, exceeding the smooth halo flux for angles $\gtrsim$ 10 degrees from the Galactic Center \citep{pieri_bertone_branchini_08,kuhlen_diemand_madau_08,springel_white_frenk_etal_08,fornasa_pieri_bertone_etal_09}.  The right panel of Figure~\ref{fig:maps} shows a realization of the signal from substructure as observed by an experiment with the angular resolution of \emph{Fermi} above $\sim 10$ GeV.

The `isotropic' diffuse gamma-ray background thus may include contributions from Galactic dark matter substructure and extragalactic dark matter, as well as from unresolved members of known extragalactic source classes such as blazars.  Since most of the measured dark matter flux may be hidden in diffuse emission, it is worthwhile to explore methods for robustly extracting this signal.  Recognizing that different source classes have different anisotropy properties, several recent studies have considered anisotropies in diffuse emission on small angular scales as a way of statistically identifying contributions from distinct source classes.  Angular power spectra of the gamma-ray emission from several extragalactic source classes \citep{ando_komatsu_06,ando_komatsu_narumoto_etal_07,miniati_koushiappas_di-matteo_07,cuoco_brandbyge_hannestad_etal_08,taoso_ando_bertone_etal_08,fornasa_pieri_bertone_etal_09} as well as from Galactic dark matter \citep{siegal-gaskins_08,fornasa_pieri_bertone_etal_09,ando_09} have been predicted.  

In this paper, I review recent theoretical work related to anisotropy analysis of gamma-ray data, however the content presented is based primarily on \citet{siegal-gaskins_08} and \citet{siegal-gaskins_pavlidou_09}.  
In Section~2 I discuss predictions for the angular power spectra of individual source classes and their relevance to the angular power spectrum of the total measured emission.  I consider the anisotropy energy spectrum, or the energy dependence of the total angular power spectrum, in Section~3 and show that the combination of anisotropy and spectral information can be a powerful tool for disentangling contributions to the diffuse emission from multiple or changing source populations.  I summarize my conclusions in Section~4.  

\section{Angular power spectra of gamma-ray source populations}
\label{sec:powerspect}

The angular power spectrum provides a statistical measure of fluctuations as a function of angular scale.  Here I consider the angular power spectrum $C_{\ell}$ of intensity fluctuations $\delta I (\psi) \equiv (I(\psi) - \langle I \rangle)/\langle I \rangle$, where $I(\psi)$ is the intensity in the direction $\psi$.  Note that the fluctuations are expressed in units of the mean map intensity, hence $\delta I$ is dimensionless.  The angular power spectrum is given by $C_{\ell} \! = \! \langle\, | a_{\ell m} |^{2} \rangle$, where $a_{\ell m}$ are determined by expanding $\delta I (\psi)$ in spherical harmonics, $\delta I (\psi)  \! = \!  \sum_{\ell,m} a_{\ell m} Y_{\ell m}(\psi)$.

For an unclustered distribution of point sources (i.e., a distribution for which the angular position of a given source is uncorrelated with the position of the other sources), the angular power spectrum is noise-like.  In this case, $C_{\ell}$ takes the same value at all $\ell$, and is inversely proportional to the number density of sources per solid angle.  Blazars are point sources and only weakly clustered, hence their angular power spectrum is well-approximated as noise-like \citep[see, e.g.,][]{ando_komatsu_narumoto_etal_07}.  Most Galactic subhalos are point-like due to the extremely small angular size of their emission region, and from our position do not appear strongly clustered, so Galactic dark matter substructure also produces a fairly noise-like power spectrum.  However, subhalos which are sufficiently massive or nearby are extended sources, and their contribution results in relatively less angular power at high multipoles (corresponding to small angular scales) compared to a pure noise-like source distribution.  In general, the amplitude of the angular power spectrum from Galactic substructure is predicted to be much greater than that from blazars.  This can be understood by noting that as a cosmological source class, blazars have a much higher number density per solid angle than Galactic substructure, and thus produce less angular power.

\begin{figure}
\includegraphics[width=0.95\textwidth]{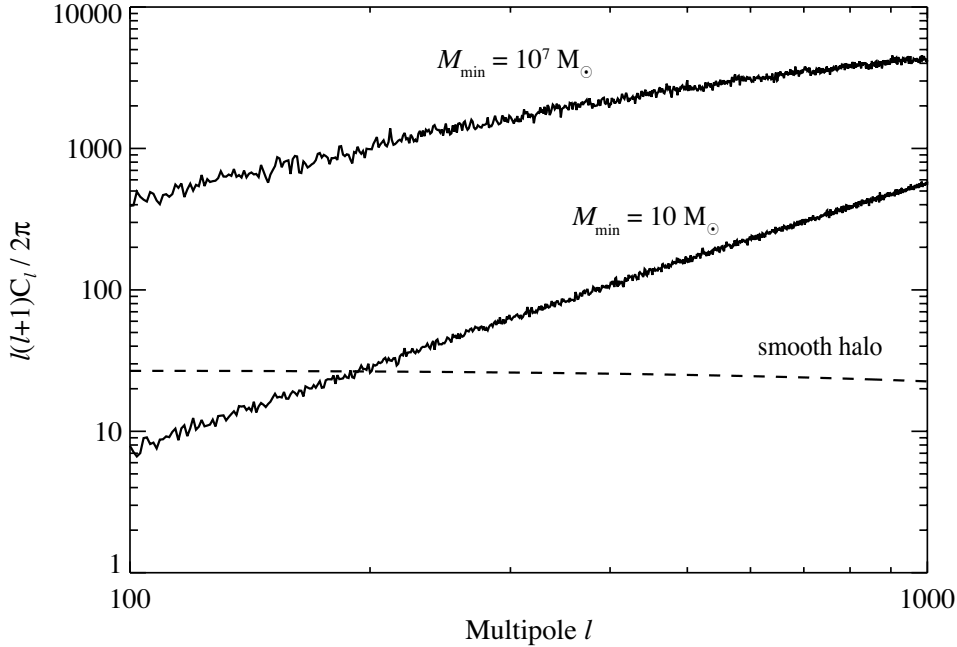}
\caption{The angular power spectrum from Galactic dark matter substructure for $M_{\rm min}=10$ M$_{\odot}$, $M_{\rm min}=10^{7}$ M$_{\odot}$, and the smooth dark matter halo.  Figure is from \citet{siegal-gaskins_08}.
\label{fig:powerspect}}
\end{figure}

Figure~\ref{fig:powerspect} shows the angular power spectrum from Galactic substructure for subhalo populations which extend to a minimum subhalo mass $M_{\rm min}=10$ M$_{\odot}$ and $M_{\rm min}=10^{7}$ M$_{\odot}$, and for the smooth dark matter halo.  In the small $M_{\rm min}$ scenario, the angular power spectrum is quite noise-like, reflecting that most of the emission is from subhalos that are effectively point sources.  For the large $M_{\rm min}$ case the angular power spectrum is flatter because more of the emission originates from extended sources.  The overall normalization of the power spectrum is higher in the $M_{\rm min}=10^{7}$ M$_{\odot}$ case since the mean map intensity is smaller, and thus the same intensity fluctuation is a larger fractional deviation from the mean than in the $M_{\rm min}=10$ M$_{\odot}$ case.  Due to its small contribution to the total intensity, the smooth halo contributes minimally to the total angular power.  It is important to keep in mind that because the angular power spectra shown are calculated in units of the mean intensity of each source class, to compare the contributions of these source classes to the total angular power spectrum the individual power spectra must be scaled by the fractional contribution of the intensity of that source class squared.

In general, the intensity and angular power spectrum of the emission from Galactic dark matter substructure exceeds that from extragalactic dark matter, and among known extragalactic source classes, unresolved blazars are the dominant contributor to both the intensity and angular power spectrum at energies $\gtrsim$ a few hundred MeV.  Here I consider only these two source classes, as they are expected to be the most relevant for analysis of the diffuse gamma-ray background with \emph{Fermi}.

\section{The anisotropy energy spectrum}
\label{sec:aniso}

In this section I present two example scenarios that demonstrate that the energy dependence of the angular power spectrum of the total measured emission could be used to confidently infer the presence of multiple contributing source populations.  In these examples I assume that the total measured isotropic diffuse emission is comprised exclusively of emission from unresolved blazars and from Galactic dark matter substructure.  For cosmological source populations such as blazars, at energies above a few tens of GeV interactions with the UV, optical, and IR backgrounds (the extragalactic background light, EBL) lead to an exponential cutoff in the observed intensity energy spectrum, so for the blazar spectra EBL attenuation is approximated here by applying an energy-dependent cut-off, and assuming for simplicity that all blazars reside at a single redshift.  The dark matter intensity energy spectrum is calculated using the approximation given in \citet{bergstrom_etal_01}.

The angular power spectrum of the total diffuse emission from extragalactic sources $C_{\ell}^{\rm EG}$ and Galactic dark matter substructure $C_{\ell}^{\rm DM}$ is given by
\begin{equation}
\label{eq:clsum}
C_{\ell}^{\rm tot}=f_{\rm EG}^{2}C_{\ell}^{\rm EG} + f_{\rm DM} ^{2}C_{\ell}^{\rm DM} + 2f_{\rm EG}f_{\rm DM}C_{\ell}^{{\rm EG}\times{\rm DM}},
\end{equation}
where $f_{\rm EG}$ and $f_{\rm DM}$ are the (energy-dependent) fractions of the total emission from extragalactic sources and Galactic dark matter substructure, respectively.  Extragalactic sources and Galactic dark matter substructure are uncorrelated, so here the cross-correlation term $C_{\ell}^{{\rm EG}\times{\rm DM}}=0$.  
As evident from Eq.~\ref{eq:clsum}, the amplitude of the total angular power spectrum at a given energy depends on the relative contributions of each source class to the measured emission at that energy.  For clarity, I use the term intensity energy spectrum to describe the energy dependence of the photon intensity.

The anisotropy energy spectrum is defined as the value of the angular power spectrum at a fixed multipole $\ell$ as a function of energy.  The angular power spectrum of a single source class in which all members have the same intensity energy spectrum is independent of energy because the angular distribution of the sources on the sky is identical at all energies.  However, contributions to the diffuse emission from multiple populations with different intensity energy spectra result in a modulation of the anisotropy energy spectrum, as the relative contribution of each population changes with energy.

\begin{figure}
\includegraphics[width=0.9\textwidth]{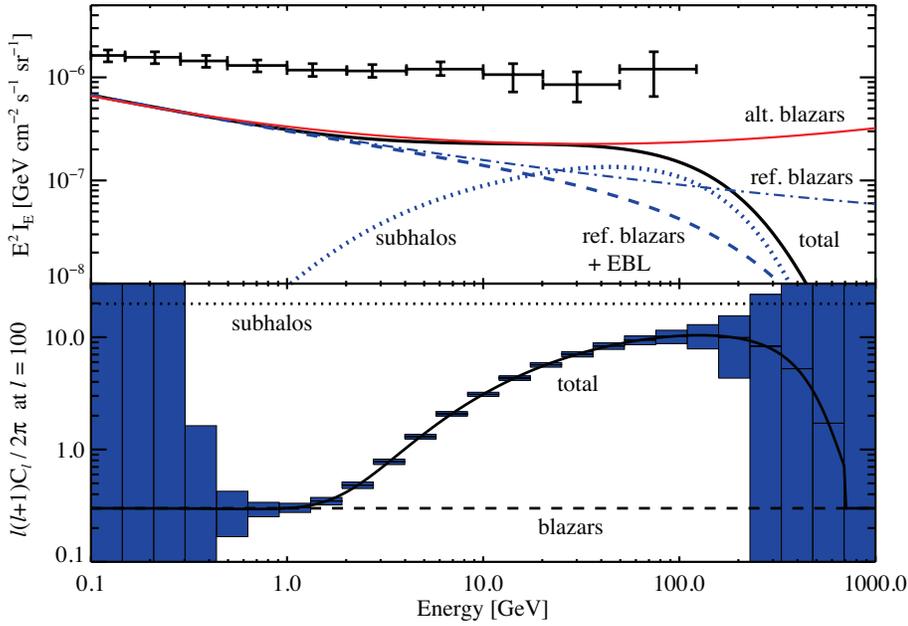}
\caption{Example intensity energy spectrum and corresponding anisotropy energy spectrum assuming a dark matter particle mass of 700 GeV.  Figure is from \citet{siegal-gaskins_pavlidou_09}.\label{fig:aniso1}}
\end{figure}

\begin{figure}
\includegraphics[width=0.9\textwidth]{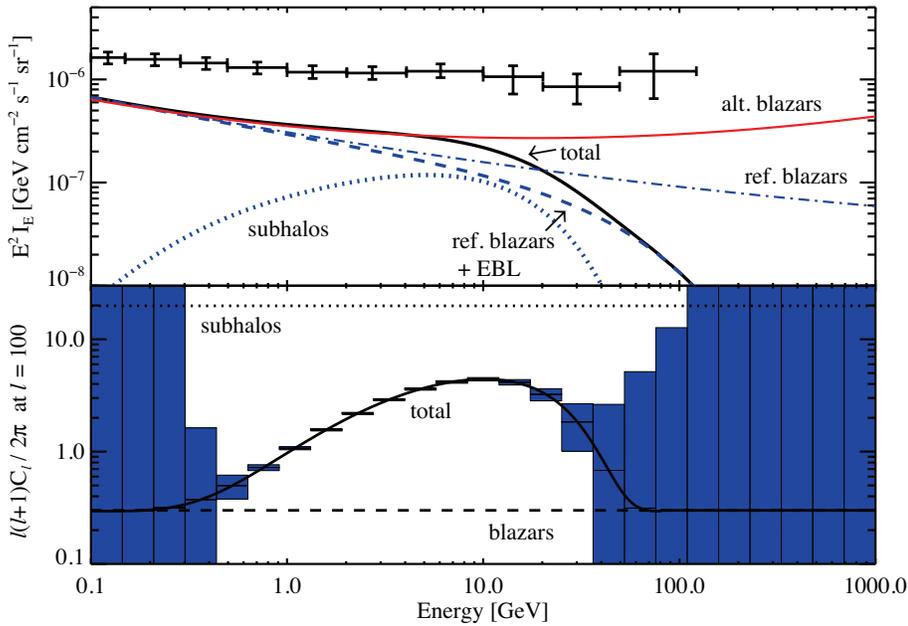}
\caption{Same as Figure~\ref{fig:aniso1}, for a dark matter particle mass of 80 GeV.  Figure is from \citet{siegal-gaskins_pavlidou_09}.\label{fig:aniso2}}
\end{figure}

Figures~\ref{fig:aniso1} and~\ref{fig:aniso2} present the intensity energy spectra and corresponding anisotropy energy spectra at $\ell=100$ for two example scenarios.  In each scenario the intensity energy spectrum of the total measured diffuse emission is consistent with that of an unresolved blazar population plus a dark matter component, but also with that of an unresolved blazar population with slightly different spectral properties without a dark matter component.  In these cases spectral information alone is insufficient to identify the dark matter signal, however, the anisotropy energy spectrum can be used to robustly identify the presence of a dark matter contribution.  In both examples the observed `total' intensity energy spectra and anisotropy energy spectra are the sum of the dark matter and the EBL-attenuated reference blazar contributions.  The intensity energy spectrum of an alternative blazar model is also shown which, with EBL attenuation, could plausibly produce the observed total intensity energy spectrum without a dark matter component.  

The scenario in Figure~\ref{fig:aniso1} assumes a dark matter mass of 700 GeV, and despite dark matter dominating the total intensity above $\sim$ 20 GeV, its contribution is not clearly identifiable in the intensity energy spectrum.  In the anisotropy energy spectrum, on the other hand, the dark matter contribution produces in a significant feature.  In Figure~\ref{fig:aniso2} the dark matter mass is assumed to be 80 GeV, and here dark matter never dominates the intensity energy spectrum, but still produces a strong feature in the anisotropy energy spectrum.  In both examples the anisotropy energy spectrum enables the blazar plus dark matter scenario to be distinguished from a blazar-only scenario: if unresolved blazars were the sole source of the isotropic diffuse emission, the anisotropy energy spectrum would be constant in energy, but the presence of a dark matter contribution results in a modulation in the anisotropy energy spectrum, unambiguously indicating a change in contributing source populations with energy.

In these figures the error bars indicate the $1-\sigma$ statistical uncertainty in the measured angular power spectrum for an all-sky observation time of 5 years with \emph{Fermi}, assuming a fraction $f_{\rm sky}=0.75$ of the sky over which the analysis is performed.
The $1-\sigma$ uncertainty is given by
\begin{equation}
\label{eq:deltacl}
\delta C_{\ell}^{\rm s} = \sqrt{\frac{2}{(2\ell + 1)\,\Delta\ell\, f_{\rm sky}}} \left(C_{\ell}^{\rm s} + \frac{C_{\rm N}}{W_{\ell}^{2}}\right),
\end{equation}
where $C_{\ell}^{\rm s}$ is the angular power spectrum of the signal (here, $C_{\ell}^{\rm tot}$) and $W_{\ell}  \! = \!  \exp(-\ell^{2}\sigma_{\rm b}^{2}/2)$ is the window function of a Gaussian beam of width $\sigma_{\rm b}$, which here corresponds to \emph{Fermi's} angular resolution.  The data is assumed to be averaged over multipole bins of width $\Delta\ell = 100$.  The noise power spectrum $C_{\rm N}$ is the sum of the Poisson noise of the signal and the background, $C_{\rm N} = (4 \pi f_{\rm sky}/N_{\rm s})(1 + (N_{\rm b}/N_{\rm s}))$, where $N_{\rm s}$ and $N_{\rm b}$ are the number of signal and background photons observed.  Unresolved blazars and Galactic dark matter substructure are considered signal, while any other observed emission, such as Galactic sources other than substructure, are noise.  The error bars shown generously assume $N_{\rm b}/N_{\rm s}=10$, i.e., an order of magnitude more background than signal photons.

For the anisotropy energy spectrum at $\ell=100$, the error bars rapidly increase at low energies ($E \lesssim 1$ GeV) due to the degradation of \emph{Fermi's} angular resolution below this energy, and at high energies due to low photon statistics.  At intermediate energies the noise term in Eq.~\ref{eq:deltacl} ($C_{\rm N}/W_{\ell}^{2}$) is quite small, and in this regime a modulation in the measured anisotropy energy spectrum can be easily identified.

It is notable that in the optimistic case that the baseline extragalactic anisotropy level is observable, it may be possible to extract the shape of the dark matter annihilation spectrum via Eq.~\ref{eq:clsum}, even in cases where the dark matter signal cannot be disentangled from other contributions in the intensity energy spectrum.

\section{Summary}
\label{sec:disc}

Anisotropies in diffuse gamma-ray emission can be a powerful diagnostic of the properties of contributing source populations.  In particular, indirect searches for dark matter may be able to use fluctuations on small angular scales to statistically detect dark matter substructure in the halo of the Galaxy even if individual subhalos are not detectable.

The robustness of anisotropy analysis can be improved by examining the energy dependence of anisotropies, since the angular power spectrum of the total measured emission varies with energy according to the energy-dependent contribution of each source population.  The examples presented in Section~3 demonstrate that a contribution from Galactic dark matter substructure to the measured diffuse emission leads to a departure of the anisotropy energy spectrum from an energy-invariant quantity.  For plausible scenarios, this modulation is shown to be detectable in \emph{Fermi} data with high confidence.

\acknowledgements 
It is a pleasure to thank my collaborator, Vasiliki Pavlidou, for her invaluable contributions to the work presented.


\begin{thebibliography}{}

\bibitem[Ando(2009)]{ando_09} Ando, S.\ 2009, arXiv:0903.4685 

\bibitem[Ando \& Komatsu(2006)]{ando_komatsu_06} Ando, S., \& Komatsu, E.\ 2006, \prd, 73, 023521 

\bibitem[Ando et al.(2007)]{ando_komatsu_narumoto_etal_07} Ando, S., Komatsu, E., Narumoto, T., \& Totani, T.\ 2007, \prd, 75, 063519 

\bibitem[Atwood et al.(2009)]{atwood_abdo_ackermann_etal_09} Atwood, W.~B., et al.\ 2009, \apj, 697, 1071 

\bibitem[Bergstr{\"o}m et al.(2001)]{bergstrom_etal_01} Bergstr{\"o}m, 
L., Edsj{\"o}, J., \& Ullio, P.\ 2001, Physical Review Letters, 87, 251301 

\bibitem[Cuoco et al.(2008)]{cuoco_brandbyge_hannestad_etal_08} Cuoco, A., Brandbyge, J., Hannestad, S., Haugb{\o}lle, T., \& Miele, G.\ 2008, \prd, 77, 123518 

\bibitem[Diemand et al.(2007)]{diemand_kuhlen_madau_07} Diemand, J., Kuhlen, M., \& Madau, P.\ 2007, \apj, 667, 859 

\bibitem[Fornasa et al.(2009)]{fornasa_pieri_bertone_etal_09} Fornasa, M., Pieri, L., Bertone, G., \& Branchini, E.\ 2009, arXiv:0901.2921 

\bibitem[Gao et al.(2004)]{gao_white_jenkins_etal_04} Gao, L., White, S.~D.~M., 
Jenkins, A., Stoehr, F., \& Springel, V.\ 2004, \mnras, 355, 819 

\bibitem[Green et al.(2005)]{green_hofmann_schwarz_05} Green, A.~M., Hofmann, S., \& Schwarz, D.~J.\ 2005, Journal of Cosmology and Astro-Particle Physics, 8, 3 

\bibitem[Kuhlen et al.(2007)]{kuhlen_diemand_madau_07} Kuhlen, M., Diemand, J., \& Madau, P.\ 2007, \apj, 671, 1135 

\bibitem[Kuhlen et al.(2008)]{kuhlen_diemand_madau_08} Kuhlen, M., Diemand, J., \& Madau, P.\ 2008, \apj, 686, 262 

\bibitem[Madau et al.(2008)]{madau_diemand_kuhlen_08} Madau, P., Diemand, J., \& Kuhlen, M.\ 2008, \apj, 679, 1260 

\bibitem[Miniati et al.(2007)]{miniati_koushiappas_di-matteo_07} Miniati, F., 
Koushiappas, S.~M., \& Di Matteo, T.\ 2007, \apjl, 667, L1 

\bibitem[Pieri et al.(2008)]{pieri_bertone_branchini_08} Pieri, L., Bertone, G., 
\& Branchini, E.\ 2008, \mnras, 384, 1627 

\bibitem[Profumo et al.(2006)]{profumo_sigurdson_kamionkowski_06} Profumo, S., Sigurdson, K., \& Kamionkowski, M.\ 2006, Physical Review Letters, 97, 031301 

\bibitem[Siegal-Gaskins(2008)]{siegal-gaskins_08} Siegal-Gaskins, J.~M.\ 
2008, Journal of Cosmology and Astro-Particle Physics, 10, 40 

\bibitem[Siegal-Gaskins \& Pavlidou(2009)]{siegal-gaskins_pavlidou_09} Siegal-Gaskins, J.~M., \& Pavlidou, V.\ 2009, arXiv:0901.3776 

\bibitem[Springel et al.(2008a)]{springel_wang_vogelsberger_etal_08} Springel, V., et al.\ 2008a, \mnras, 391, 1685 

\bibitem[Springel et al.(2008b)]{springel_white_frenk_etal_08} Springel, V., et al.\ 
2008b, \nat, 456, 73 

\bibitem[Taoso et al.(2009)]{taoso_ando_bertone_etal_08} Taoso, M., Ando, S., 
Bertone, G., \& Profumo, S.\ 2009, \prd, 79, 043521 


\end{thebibliography}
\end{document}